\documentclass[floatfix,aps,twocolumn,superscriptaddress,prb]{revtex4-2}

\usepackage{graphicx}% Include figure files
\usepackage{dcolumn}% Align table columns on decimal point
\usepackage{bm}% bold math
\usepackage[T1]{fontenc}
\usepackage[latin9]{inputenc}
\usepackage{textcomp}
\usepackage{amsmath}
\usepackage{mathrsfs}
\usepackage{graphicx}
\usepackage{amssymb}
\usepackage{units}
\usepackage{xcolor}
\usepackage{float}
\usepackage{changes}

\begin{document}

\title{Polaron formation in insulators
and the key role of hole scattering processes
: \\ Band insulators, charge density waves and Mott transition}

\author{Ivan Amelio}
\affiliation{Center for Nonlinear Phenomena and Complex Systems,
Universit{\'e} Libre de Bruxelles, CP 231, Campus Plaine, B-1050 Brussels, Belgium}
\author{Giacomo Mazza}
\affiliation{Department of Physics ``E. Fermi'' University of Pisa, Largo B. Pontecorvo 3, 56127 Pisa, Italy}
\author{Nathan Goldman}
\affiliation{Center for Nonlinear Phenomena and Complex Systems,
Universit{\'e} Libre de Bruxelles, CP 231, Campus Plaine, B-1050 Brussels, Belgium}
\affiliation{Laboratoire Kastler Brossel, Coll\`ege de France, CNRS, ENS-PSL University, Sorbonne Universit\'e, 11 Place Marcelin Berthelot, 75005 Paris, France
}

\begin{abstract}
A mobile impurity immersed in a non-interacting Fermi sea is dressed by the gapless particle-hole excitations of the fermionic medium. This conventional Fermi-polaron setting is well described by the so-called ladder approximation, which consists in neglecting impurity-hole scattering processes.
In this work,  we analyze polaron formation in the context of insulating states of matter, considering increasing levels of correlation in the medium:~band insulators originating from external periodic potentials, spontaneously-formed charge density waves, and a Fermi-Hubbard system undergoing a metal-Mott insulator transition. The polaron spectral function is shown to exhibit striking signatures of the underlying fermionic background, such as the single-particle band gap, particle-hole symmetry and the transition to the Mott state. These signatures are identified within the framework of the Chevy ansatz, i.e.~upon restricting the Hilbert space to single particle-hole excitations.
Interestingly, we find that the ladder approximation is inaccurate in these band systems, due to the fact that the particle and hole scattering phase spaces are comparable.
Our results provide a step forward in the understanding of polaron formation in correlated many-body media, which are relevant to both cold-atom and semiconductor experiments.

\end{abstract}

\date{\today}
%\jobname

\maketitle

\section{Introduction}

Immersing an impurity in a many-body bath  provides a paradigmatic theoretical and experimental setting, in which the dressing of the impurity by the bath excitations gives rise to a new object, conventionally called  {\em polaron}.  

In recent years, 
there has been great interest in the physics of polarons due to the possibility of engineering mixtures of ultracold atoms, as well as 
of probing the states of  few-layer transition metal dichalchogenides (TMD) heterostructures via polaron spectroscopy.
A few ultracold atom experiments have been able to access the spectral features of Fermi and Bose polarons, exploiting the  tunability of two-body interactions via Feschbach resonances~\cite{Schirotzek_2009, Nascimbene_2009, Kohstall_2012, Koschorreck_2012, hu2016, jorgensen2016, yan2020}.
In the semiconductor context, 
optical spectroscopy is sensitive to the dressing of the excitons by the electronic excitations of the two-dimensional material; for example, trion resonances emerge when doping a monolayer~\cite{sidler2017fermi,courtade2017charged}.

In the simplest situation, the bath consists of a non-interacting Fermi sea in the continuum. 
An extremely convenient description of Fermi polaron physics is provided by the so-called (non-self-consistent) T-matrix approximation, or ladder approximation, where the fermion-impurity T-matrix is computed exactly and the self-energy of the impurity is obtained by summing over all possible hole states. 
In this framework, the hole-impurity scattering processes are neglected, which is justified by the fact that the phase space for particle scattering is infinitely larger than the one for holes.
It is easily proven that, for contact interactions and a continuum system, the T-matrix approximation coincides with the variational treatment proposed by Chevy~\cite{chevy2006universal}, where the Hilbert space of the fermions is restricted to include at most one particle-hole excitation on top of the Fermi sea.
The reliability of the T-matrix
approach to the Fermi polaron
has been confirmed by a number of numerical benchmarks~\cite{prokofev2008bold,combescot2008,vanhoucke2020}.

Conversely, only a relatively small  number of theoretical works have recently investigated, using different techniques, polaron formation in more complex and correlated baths,
including
interacting Bose gases~\cite{schmidt2022self,sanchezbaena2023universal} and Bose-Hubbard lattices~\cite{colussi2023lattice,santiagogarcia2024lattice},
fermionic superfluids~\cite{yi2015polarons,amelio2023two-dimensional},
correlated electrons in twisted heterostructure~\cite{mazza2022strongly},
excitonic insulators~\cite{amelio2023polaron},
charge density waves~\cite{amelio2024edpolaron},
supersolids~\cite{simons2024polaronsupersolid},
topological systems~\cite{Grusdt2016,grusdt2019,munoz2020anyonic,vashisht2024chiral}
and unconventional single-particle bands~\cite{Sorout_2020,pimenov2024}.
On the experimental side, TMD heterostructures have allowed to access  non-trivial spectral features in the context of (generalized) Wigner crystals~\cite{Regan2020mott,Xu2020correlated,smolenski2021signatures, Zhou2021bilayer, Li2021imaging, shimazaki2021optical,kiper2024confinedtrionsmottwignerstates}, correlated insulators~\cite{Xu2020correlated,shimazaki2020strongly},
quantum Hall liquids~\cite{smolenski2019interaction-induced,Popert2022,Cai2023} and kinetic magnetism~\cite{ciorciaro2023, tao2024}.

This work sets the focus on the formation of polarons in insulating states of fermionic matter. As prime examples, we consider band insulators, charge density waves, as well as the Mott transition occurring in the strongly-interacting Fermi-Hubbard model. 
Our study reveals the prominent roles of translational-symmetry breaking, the presence of a single-particle gap and the existence of particle-hole symmetry in this context, which all lead to characteristic signatures in the polaron spectrum.

On a more technical side, to what extent the T-matrix and Chevy approximations hold in these more complicated scenarios remains an open question. 
In the following, we take a small step forward and  demonstrate that, in general, 
the T-matrix and Chevy approximation can yield significantly different results.
In particular, hole scattering processes cannot be neglected in few band systems or at half-filling, where the phase space for particle and hole scattering are comparable.
This situation is relevant to 
optical lattices and moir\'e systems, provided that the impurity-fermion interactions are small enough to be able to consider a few bands.

The paper is structured as follows. 
We first review in Sec. \ref{sec:reminder} the ladder approach to the polaron in a non-interacting Fermi sea, and its relation to the Chevy approximation.
In Sec. \ref{sec:cdw}, we apply Bloch theorem to the study of polaron spectra in a background with a periodic fermionic density, 
arising either from an external potential or a spontaneously formed charge density wave (CDW),
 computed using Hartree-Fock theory. A very characteristic avoided crossing in the repulsive polaron branch occurs when the two-body binding energy 
is comparable with the inter-band gap.
Attractive polaron binding energies are overestimated and spurious peaks are obtained when neglecting hole scattering processes.
In Sec. \ref{sec:mott},
we address the polaron formation in the case in which the underlying fermionic bath undergoes a Mott transition.
We assume for simplicity an infinite mass for the impurity and the Chevy approximation, and observe a cusp at the metal-insulator transition. The role of particle-hole symmetry of the bath is elucidated.
Finally, we draw our conclusions and discuss perspectives in Sec. \ref{sec:conclusions}.

\section{Reminder of the T-matrix and Chevy ansatz approximations}
\label{sec:reminder}

In this Section, we briefly review the (non-self-consistent) T-matrix and Chevy approaches to the Fermi polaron in a non-interacting Fermi sea defined on a 2D continuum plane.

The Hamiltonian of the system reads
\begin{equation}
    H =
    \sum_{\mathbf{k}}
    \xi_{\mathbf{k}} 
    c^\dagger_{\mathbf{k}} c_{\mathbf{k}}
    +
    \sum_{\mathbf{q}}
    \epsilon^X_{\mathbf{q}} 
    x^\dagger_{\mathbf{q}} x_{\mathbf{q}}
    +
    \frac{g}{\mathcal{V}}
    \sum_{\mathbf{k}\mathbf{p}\mathbf{q}}
     c^\dagger_{\mathbf{k}} c_{\mathbf{p}}
    x^\dagger_{\mathbf{q}+\mathbf{p}-\mathbf{k}} x_{\mathbf{q}},
\end{equation}
where 
$c^\dagger_{\mathbf{k}}$
creates a spinless fermion at momentum ${\mathbf{k}}$
in a 2D box of area $\mathcal{V}$, while
$x^\dagger_{\mathbf{q}}$ is the creation operator of the impurity (the letter $x$ reminds of  the exciton, used as impurity in optical spectroscopy in TMDs).
The bare dispersion of the fermion is 
$\xi_{\mathbf{k}} = \frac{\mathbf{k}^2}{2m} - \mu$
with $m$ the mass and $\mu$ the Fermi level,  the impurity instead has mass $M$ and dispersion 
$\epsilon^X_{\mathbf{q}}
=\epsilon^X + \frac{\mathbf{q}^2}{2M}
$ with $\epsilon^X$ the bottom of the exciton band,
while the impurity-fermion contact interaction has strength $g$.

For attractive interactions $g<0$, the 
Lippmann-Schwinger equation 
\begin{equation}
    \frac{1}{g}
    =
    \frac{1}{\mathcal{V}}
    \sum^{k_\infty}_{\mathbf{k}}
    \frac{1}{-E_B-\frac{\mathbf{k}^2}{2m_{\rm red}}},
\end{equation}
yields the binding energy $E_B$
between a fermion and an impurity in vacuum (here
$m_{\rm red}=mM/(m+M)$ denotes the reduced mass).
The sum has a logarithmic divergence and needs to be regularized by introducing the cutoff momentum $k_\infty$.
Typically, one assumes that $E_B$ is fixed (e.g. measured experimentally) and solves the theory sending $k_\infty$ to infinity.
When the interaction is renormalized in this way, $g$ goes to zero.

\begin{figure}[t]
    \centering
\includegraphics[width=0.46\textwidth]{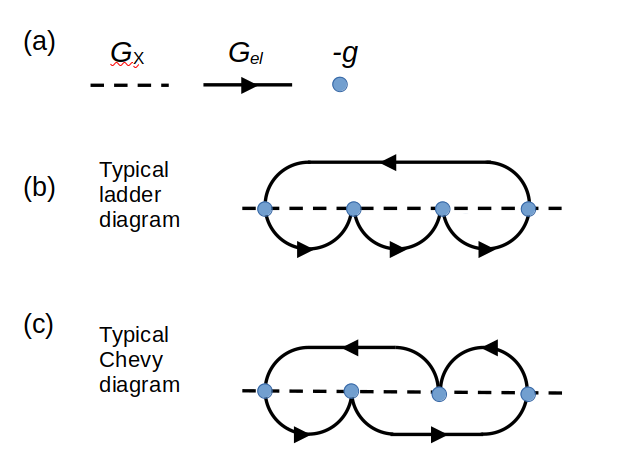}
    \caption{
    Example of typical Feynmann diagrams. (a) Impurity propagator $G_X$ (dashed line), fermionic propagator $G_{el}$ (solid line with rightward arrow), and bare vertex (blue circle).
    The hole propagator corresponds to the fermionic propagator with leftward arrow.
    (b) A typical impurity-particle diagram participating in the standard ladder resummation.
    (c) A diagram including also impurity-hole scattering and neglected in the T-matrix approach, but in principle contributing in the Chevy ansatz formalism.}
    \label{fig:simple_diagrams}
\end{figure}

A simple but effective approach to the physics of the polaron,
is to truncate the Hilbert space to include up to one single particle-hole pair excitation on top of the non-interacting Fermi sea $|\mathcal{F}\rangle$.
A generic state of total momentum $\mathbf{q}$ (which is a conserved quantity) then takes  the Chevy ansatz \cite{chevy2006universal} form
\begin{equation}
    |\Psi_\mathbf{q} \rangle 
    =
    \left\{ \psi^\mathbf{q} x_\mathbf{q}^\dagger +
    \sum_{\mathbf{k}\mathbf{p}} \psi^\mathbf{q}_{\mathbf{k}\mathbf{p}} c^\dagger_\mathbf{k} c_\mathbf{p}
      x_{\mathbf{q}-\mathbf{k}+\mathbf{p}}^\dagger 
      \right\}
      |\mathcal{F}\rangle,
\end{equation}
where in the summation it is implicit that $\mathbf{k},\mathbf{p}$ run over the empty and occupied fermionic states, respectively.

In this basis, the Schr\"odinger equation 
translates into the equations of motion for the wavefuntion entries $\psi^\mathbf{q}_{\mathbf{k}\mathbf{p}}, \psi^\mathbf{q}$, namely
\begin{equation}
    i\partial_t \psi^\mathbf{q}
    =
    (\epsilon^X_{\mathbf{q}} + gn) \psi^\mathbf{q}
    + \frac{g}{\mathcal{V}} 
    \sum_{\mathbf{k}\mathbf{p}} 
    \psi^\mathbf{q}_{\mathbf{k}\mathbf{p}},
    \label{eq:simpleChevy1}
\end{equation}
\begin{multline}
    i\partial_t \psi^\mathbf{q}_{\mathbf{k}\mathbf{p}}
    =
    (\epsilon^X_{\mathbf{q}-\mathbf{k}+\mathbf{p}} 
    + \xi_{\mathbf{k}}
    - \xi_{\mathbf{p}}
    +gn) \psi^\mathbf{q}_{\mathbf{k}\mathbf{p}}
    +
    \\
    +
    \frac{g}{\mathcal{V}}  
    \psi^\mathbf{q}
    +
    \frac{g}{\mathcal{V}}
    \sum_{\mathbf{k}'}
    \psi^\mathbf{q}_{\mathbf{k}'\mathbf{p}}
    -
    \frac{g}{\mathcal{V}}
    \sum_{\mathbf{p}'}
    \psi^\mathbf{q}_{\mathbf{k}\mathbf{p}'},
        \label{eq:simpleChevy2}
\end{multline}
with $n$ the fermion density.
The last term of Eq.~(\ref{eq:simpleChevy1}) describes the creation of a particle-hole pair by scattering with the impurity, 
while the last two terms of Eq.~(\ref{eq:simpleChevy1})
correspond, respectively, to impurity-particle and impurity-hole scattering processes.

A crucial remark is that, when the interaction is renormalized sending $g \to 0$,
the mean-field shift $gn$ and the hole-scattering term are negligible, due to the finite  hole phase space proportional to the Fermi energy. The electron scattering term is instead relevant, because the particle phase space is unbounded.

Dropping the hole term allows to resum Eqs.~(\ref{eq:simpleChevy1},\ref{eq:simpleChevy2}) and yields the self-consistent condition
$E = \epsilon^X + \Sigma_X(\mathbf{q}, E)$
for the energy eigenvalues $E$~\footnote{
With the renormalization $g \to 0$ and introducing
the auxiliary quantity $\chi^\mathbf{q}_{\mathbf{p}} = \frac{g}{\mathcal{V}}(\psi^\mathbf{q} + \sum_\mathbf{k} \psi^\mathbf{q}_{\mathbf{k}\mathbf{p}})$, Eqs.~(\ref{eq:simpleChevy1},\ref{eq:simpleChevy2}) are recast into
$$
\psi^\mathbf{q} = \frac{1}{E-\epsilon^X_\mathbf{q}} \sum_\mathbf{p} \chi^\mathbf{q}_{\mathbf{p}},
$$
$$
\psi^\mathbf{q}_{\mathbf{k}\mathbf{p}}
=
\frac{1}{E-\frac{\mathbf{k}^2}{2m}-\epsilon^X_{\mathbf{q}-\mathbf{k}}} \chi^\mathbf{q}_{\mathbf{p}},
$$
which can be reinserted into the definition of $\chi^\mathbf{q}_{\mathbf{p}}$. Rearranging and summing over $\mathbf{p}$ finally yields
$(E-\epsilon^X_\mathbf{q}) \sum_\mathbf{p} \chi^\mathbf{q}_{\mathbf{p}} = \Sigma_X(\mathbf{q}, E) \sum_\mathbf{p} \chi^\mathbf{q}_{\mathbf{p}}$
}.
Here we introduced
the impurity self-energy
\begin{equation}
    \Sigma_X(\mathbf{q}, \omega)
    = 
    % g n + 
    \frac{1}{\mathcal{V}}
    \sum_{|\mathbf{p}|<k_F} 
   % n_F(\xi_{\mathbf{p}}) 
     \Tilde{\Lambda}(\mathbf{q} + \mathbf{p},  \omega + \xi_{\mathbf{p}}),
    \label{eq:SigmaX_simple}
\end{equation}
where the two-body T-matrix 
\begin{equation}
    \Tilde{\Lambda}(\mathbf{q},  \omega)
    =
    \left[ 
    \frac{1}{g}
    -
    \frac{1}{\mathcal{V}}
    \sum_{|\mathbf{k}|>k_F}
    \frac{1}{\omega-\frac{\mathbf{k}^2}{2m}-\epsilon^X_{\mathbf{q}-\mathbf{k}}}
    \right]^{-1}
    \label{eq:tilde_Lambda}
\end{equation}
obeys the Dyson equation
$\Tilde{\Lambda} = g + g \Pi_{\rm X-el} \Tilde{\Lambda}$ with $\Pi_{\rm X-el} = \frac{1}{\mathcal{V}}
    \sum_{|\mathbf{k}|>k_F}
    [\omega-\frac{\mathbf{k}^2}{2m}-\epsilon^X_{\mathbf{q}-\mathbf{k}}]^{-1}$,
and
corresponds to the resummation of the ladder diagrams
describing particle-impurity scattering.
Equations. (\ref{eq:SigmaX_simple},\ref{eq:tilde_Lambda}) constitute the so-called non-self-consistent T-matrix approximation, aka ladder approximation.
So far, we have shown that, in continuum systems with contact interactions, the Chevy approach and non-self-consistent T-matrix method are equivalent, as it is well known~\cite{combescot2007normal,parish2011polaron-molecule,schmidt2012fermi,parish2023fermi}.
One of the main goals of this paper is to study situations where the particle and hole phase spaces have comparable extent, and this equivalence does not hold.

Some diagrammatic intuition can  be developed by inspecting
Fig.~\ref{fig:simple_diagrams}. In panel (a) we define the bare propagators and vertex, while in panels (b) and (c) we report typical sample diagrams corresponding, respectively, to  T-matrix and Chevy contributions to the impurity self-energy.
Notice that the restriction to the Chevy subspace entails that at each time at most one particle-hole pair can be drawn.
A fermionic solid line with arrow pointing leftwards is interpreted as a hole propagator. 
In particular, in Fig.~\ref{fig:simple_diagrams}.(b) only impurity-particle scattering processes are present, and can be easily resummed using the ladder method, to yield Eqs. (\ref{eq:SigmaX_simple},\ref{eq:tilde_Lambda}). 
Conversely, the Chevy approach depicted in panel (c) includes impurity-hole scattering.
We will show in  Appendix \ref{app:A} how   a resummation is formally possible in this case as well.

\section{Polarons in charge density waves and band insulators}
\label{sec:cdw}

In this Section we will investigate polaron spectra on top of band insulators and CDWs.
In band insulators, a  periodic external potential creates a set of Bloch bands and the Fermi level lies within a gap between two sets of these bands.
At the Chevy level, it is assumed that a hole is created in one of the valence bands and a fermion is promoted to the set of conduction bands.
The Chevy equations of motion are written down explicitly  keeping track of the discrete translational invariance of the system, and are labeled by the Bloch quasi-momentum of the polaron.

Then, we consider charge density waves in the Hartree-Fock approximation. Within the Hartree-Fock method, the fermions are quasi-particles occupying 
effective single particle bands generated by the non-local Hartree-Fock potential~\cite{Giuliani2005}. As a consequence, the polaron equations developed for band insulators apply straightforwardly. 

We remark that in this work we restrict numerical results to two-dimensional lattice systems, meaning that the periodic external potentials or the spontaneously formed CDWs lie on top of a microscopic lattice.
This choice is motivated both by technical reasons (smaller matrices to diagonalize, no ultraviolet regularization needed) and experimental considerations in moir\'e systems, since CDWs and generalized Wigner crystals are 
favored by the presence of the moir\'e potential~\cite{Regan2020mott,Xu2020correlated,Li2021imaging}.
Our formalism and main results should extend to other dimensionalities and continuum systems.

\subsection{Chevy formalism for  Bloch bands}

Here, we consider here a system of spinless fermions in a periodic external potential $V^{\rm ext}(\mathbf{r})$. The Hamiltonian reads in momentum space
\begin{equation}
    H^{\rm el} = \sum_{\Tilde{\mathbf{k}}} \epsilon_{\Tilde{\mathbf{k}}} c_{\Tilde{\mathbf{k}}}^\dagger c_{\Tilde{\mathbf{k}}}
    +
    \frac{1}{\mathcal{V}}\sum_{\mathbf{k}\mathbf{G}\mathbf{G}'}
    V^{\rm ext}_{\mathbf{G}-\mathbf{G}'}
    c_{\mathbf{k}+\mathbf{G}}^\dagger c_{\mathbf{k}+\mathbf{G}'},
\end{equation}
where $\epsilon_{\Tilde{\mathbf{k}}}$ is the bare dispersion of the fermions, $c^\dagger_{\Tilde{\mathbf{k}}}$ creates a fermion with momentum $\Tilde{\mathbf{k}} = \mathbf{k}+\mathbf{G}$, where $\mathbf{k}$ is the Bloch quasi-momentum within the first Brillouin zone (1BZ) and $\mathbf{G}$ belongs to the reciprocal lattice of the periodic potential. This notational convention is implicit in the summations and in the following. Finally, $\mathcal{V}$ denotes the total area or volume for a continuous system, or the total number of lattice sites in a discretized setting. 

The Bloch theorem ensures that the Hamiltonian can be block-diagonalized within each quasi-momentum sector, yielding
\begin{equation}
    H^{\rm el} 
    =
    \sum_{\mathbf{k}\alpha}
    E_{\mathbf{k}\alpha} 
    d^\dagger_{\mathbf{k}\alpha} d_{\mathbf{k}\alpha},
\end{equation}
with $d^\dagger_{\mathbf{k}\alpha} = \sum_{\mathbf{G}} u_{\mathbf{k}\alpha}(\mathbf{G}) \ c^\dagger_{\mathbf{k}+\mathbf{G}}$
the quasi-particle creation operator,
where $\alpha$ is the band index, $E_{\mathbf{k}\alpha}$ the band energy at a given quasi-momentum, and  $u_{\mathbf{k}\alpha}(\mathbf{G})$ the periodic Bloch wavefunctions.

In the following, we assume that the fermion-impurity interaction is a contact one with coupling $g$,
\begin{equation}
    H^{\rm el- X} =
    \frac{g}{\mathcal{V}}\sum_{\Tilde{\mathbf{k}}\Tilde{\mathbf{p}}\Tilde{\mathbf{q}}}
    c_{\Tilde{\mathbf{k}}}^\dagger c_{\Tilde{\mathbf{p}}}
    x_{\Tilde{\mathbf{q}}-\Tilde{\mathbf{k}}+\Tilde{\mathbf{p}}}^\dagger x_{\Tilde{\mathbf{q}}}.
\end{equation}

The fermion-impurity binding energy $E_B$ in vacuum and in the absence of external potential is given by the 
Lippmann-Schwinger equation 
\begin{equation}
    \frac{1}{g}
    =
    \frac{1}{\mathcal{V}}
    \sum_{\Tilde{\mathbf{k}}}
    \frac{1}{-E_B-\epsilon^f_{\Tilde{\mathbf{k}}}-
    \epsilon^X_{-\Tilde{\mathbf{k}}}},
\end{equation}
and in two dimensions $E_B>0$ for any $g<0$. 
In the continuum, an ultraviolet cutoff, corresponding to the physical range of the interaction, is needed to make the sum convergent; in particular, $g$ tends to zero as the cutoff is sent to infinity.
On a lattice, instead, the Bravais unit length provides a natural regularization.
In a moir\'e system, one can imagine that only the lowest mini-moir\'e band is active, with the range of the exciton-electron interaction being a few times smaller than the separation between the moir\'e minima. 
A superpotential $V^{\rm ext}(\mathbf{r})$, with larger period, can then be superimposed, or a CDW  can form within the first mini-band.

\begin{figure*}[t]
    \centering
\includegraphics[width=0.90\textwidth]{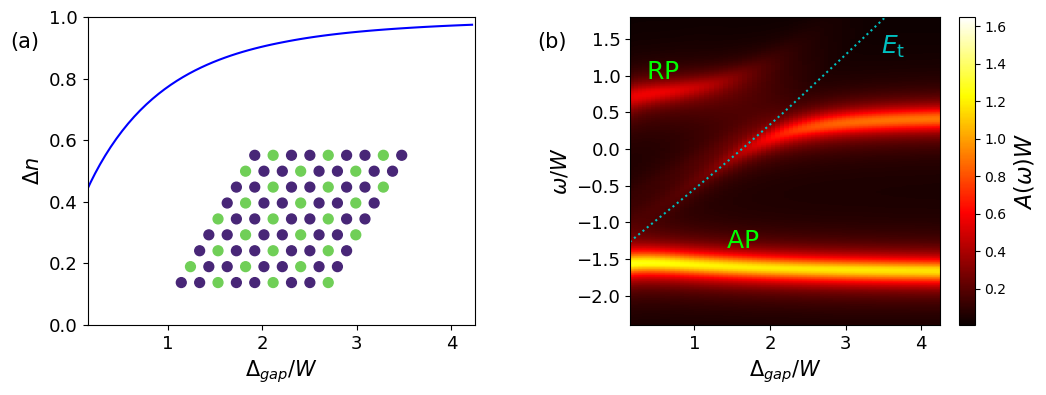}
    \caption{
      Chevy computation for a 9$\times$9 triangular lattice at one-third filling, the CDW being described at the Hartree-Fock level.
    (left)
    Density contrast of the charge density wave $\Delta n$ as a function of the quasi-particle gap $\Delta_{\rm gap}$ (in units of the bandwidth $W$), scanned by varying the fermion repulsion $\mathcal{Z}$.
    Inset: the real space CDW density pattern.
    (right)
    Polaron spectra for $E_B/W = 0.8$. Notice the renormalization of the  AP due to interactions and the avoided crossing in the RP occurring for a gap of the order of $|E_{AP}|$. }
    \label{fig:ChevyHF}
\end{figure*}

At zero temperature, if the Fermi level falls within a bandgap, then the fermionic ground state --in the absence of the impurity-- consists of $N_{\rm v}$
filled valence bands.
We denote this insulating ground state as 
$| \mathcal{F }
\rangle = \Pi_{\mathbf{k}\beta} d^\dagger_{\mathbf{k}\beta} |0\rangle$,
where $\mathbf{k} \in 1BZ, \beta=1,...,N_{\rm v}$.
When an impurity is created in such a many-body background (by the operator $x^\dagger_{\mathbf{q}}$),
it will excite particle-hole pairs  out of $|\mathcal{F }\rangle$.
If one restricts to up to one such excitation,
the state of the system in the total quasi-momentum $\mathbf{q}$ takes the 
Chevy ansatz form
\begin{multline}
    | \Psi_\mathbf{q} \rangle
    =
    \left\{
\sum_\mathbf{G} \psi_{\mathbf{G}} x^\dagger_{\mathbf{q}+\mathbf{G}} 
+ \right.
\\
+
\left.
\sum_{\mathbf{k}\mathbf{p}\mathbf{G}\alpha\beta} 
\psi_{\mathbf{k}\mathbf{p}}^{\alpha \beta \mathbf{G}} \
d^\dagger_{\mathbf{k}\alpha} d_{\mathbf{p}\beta}
x_{\mathbf{q}+\mathbf{p}-\mathbf{k}+\mathbf{G}}^\dagger
    \right\} | \mathcal{F }\rangle,
\end{multline}
where $\mathbf{k},\mathbf{p},\mathbf{q} \in 1BZ, \beta=1,...,N_{\rm v}$ and $\alpha>N_{\rm v}$ run over the valence and conduction band indices, respectively, and $\mathbf{G}$ belongs to the  reciprocal lattice of $V^{\rm ext}$. Here, $\psi_{\mathbf{G}}, \psi_{\mathbf{k}\mathbf{p}}^{\alpha \beta \mathbf{G}}$
are just the coefficients of the basis expansion in this sector of the Hilbert space.
(There is no physical meaning for upper versus lower indices, it is just a graphics choice.)
Notice that, in the second summation, $\alpha$ and $\beta$ indices run over the empty and filled bands, respectively.

Denoting as $H^{\rm X} = \sum_{\Tilde{\mathbf{q}}} \epsilon^X_{\Tilde{\mathbf{q}}}
x^\dagger_{\Tilde{\mathbf{q}}} x_{\Tilde{\mathbf{q}}}$ the bare kinetic energy operator of the impurity, the total Hamiltonian of the system reads
\begin{equation}
    H = H^{\rm el} + H^{\rm X}
    +
    H^{\rm el-X}.
\end{equation}
The equations of motion for $ | \Psi_\mathbf{q} \rangle $ then read
\begin{multline}
    i\partial_t \psi_\mathbf{G} 
    =
    \epsilon^X_{\mathbf{q}+\mathbf{G}} \psi_\mathbf{G}
    +
     \sum_{\mathbf{G}'} 
   (\mathcal{M}^{\rm 0})_{\mathbf{G}\mathbf{G}'} \
\psi_{\mathbf{G}'}
    +
    \\
+
    \sum_{\mathbf{k}\mathbf{p}\mathbf{G}'\alpha\beta} 
   (\mathcal{M}^{\rm ph})^{\alpha\beta \mathbf{G}\mathbf{G}'}_{\mathbf{k}\mathbf{p}}  \
\psi_{\mathbf{k}\mathbf{p}}^{\alpha\beta \mathbf{G}'}
\label{eq:Chevy_CDW_1}
\end{multline}
\begin{multline}
    i\partial_t \psi_{\mathbf{k}\mathbf{p}}^{\alpha\beta \mathbf{G}} 
    =
    (\epsilon^X_{\mathbf{q}+\mathbf{p}-\mathbf{k}+\mathbf{G}} + E_{\mathbf{k}\alpha} - E_{\mathbf{p}\beta}) \psi_{\mathbf{k}\mathbf{p}}^{\alpha\beta \mathbf{G}}
    +
    \\
+
    \sum_{\mathbf{G}'} 
(\mathcal{M}^{\rm 0})_{\mathbf{G}\mathbf{G}'}  \
\psi_{\mathbf{k}\mathbf{p}}^{\alpha \beta G'}
+
    \sum_{\mathbf{G}'} 
   (\mathcal{M}^{\rm ph})^{\alpha\beta \mathbf{G}\mathbf{G}'}_{\mathbf{k}\mathbf{p}} \
\psi_{\mathbf{G}'}
+
\\
+
 \sum_{\mathbf{k}'\mathbf{G}'\alpha'} 
(\mathcal{M}^{\rm p})^{\alpha\alpha'\beta \mathbf{G}\mathbf{G}'}_{\mathbf{k}\mathbf{k}'\mathbf{p}} \
\psi_{\mathbf{k}'\mathbf{p}}^{\alpha'\beta \mathbf{G}'}
-
 \sum_{\mathbf{p}'\mathbf{G}'\beta'} 
(\mathcal{M}^{\rm h})^{\alpha\beta\beta' \mathbf{G}\mathbf{G}'}_{\mathbf{k}\mathbf{p}\mathbf{p}'} \
\psi_{\mathbf{k}\mathbf{p}'}^{\alpha\beta' \mathbf{G}'},
\label{eq:Chevy_CDW_2}
\end{multline}
where we will now analyze individually each term.
First of all, the diagonal terms contain the bare impurity and quasi-particle dispersion.
Then, the mean-field scattering term is
\begin{equation}
    (\mathcal{M}^{\rm 0})_{\mathbf{G}\mathbf{G}'}
    =
    \frac{g}{\mathcal{V}} \sum_{\mathbf{k}\mathbf{R}\beta} 
   u^*_{\mathbf{k}\beta}(\mathbf{R}) u_{\mathbf{k}\beta}(\mathbf{G}+\mathbf{R}-\mathbf{G}'),
\end{equation}
and it describes the scattering of the impurity by the static density profile of the ground state.
In particular, this term is present also in the subspace with zero particle-hole excitations.

Moving to the
1-particle-hole sector, the
particle scattering term 
\begin{equation}
    (\mathcal{M}^{\rm p})^{\alpha\alpha'\beta \mathbf{G}\mathbf{G}'}_{\mathbf{k}\mathbf{k}'\mathbf{p}}
    =
    \frac{g}{\mathcal{V}} \sum_{\mathbf{R}} 
u^*_{\mathbf{k}\alpha}(\mathbf{R}) u_{\mathbf{k}'\alpha'}(\mathbf{G}+\mathbf{R}-\mathbf{G}')
\end{equation}
describes scattering of the fermion by the impurity within the conduction band phase space.
Similarly,
\begin{equation}
    (\mathcal{M}^{\rm h})^{\alpha\beta\beta' \mathbf{G}\mathbf{G}'}_{\mathbf{k}\mathbf{p}\mathbf{p}'}
    =
    \frac{g}{\mathcal{V}} \sum_{\mathbf{R}} 
u^*_{\mathbf{p}'\beta'}(\mathbf{R}) u_{\mathbf{p}\beta}(\mathbf{G}+\mathbf{R}-\mathbf{G}')
\end{equation}
describes impurity-hole scattering within the valence band states.
Finally, the particle-hole generation term 
\begin{equation}
    (\mathcal{M}^{\rm ph})^{\alpha\beta \mathbf{G}\mathbf{G}'}_{\mathbf{k}\mathbf{p}}
    =
    \frac{g}{\mathcal{V}} \sum_{\mathbf{R}\mathbf{G}'} 
   u^*_{\mathbf{p}\beta}(\mathbf{R}) u_{\mathbf{k}\alpha}(\mathbf{G}+\mathbf{R}-\mathbf{G}') \ 
\end{equation}
connects the subspace with zero and one particle-hole excitations.

The polaron spectral function is given by
\begin{multline}
    A_{\rm X}(\mathbf{q},\omega)
\equiv
- \frac{1}{\pi}
{\rm Im}
\langle  \mathcal{F } |
x_{\mathbf{q}}
\frac{1}{\omega - H + i0^+}
x_{\mathbf{q}}^\dagger
|  \mathcal{F } \rangle
=
\\
= - \frac{1}{\pi}
{\rm Im}
\sum_n \frac{|\psi^{(n)}_{\mathbf{q}}|^2}{\omega - E_n + i0^+},
\label{eq:A_X}
\end{multline}
where the sum is over the eigenstates
$\psi^{(n)}$ and eigenergies $E_n$ of Eqs. (\ref{eq:Chevy_CDW_1},\ref{eq:Chevy_CDW_2}).
The spectral weight is given by the component of the Chevy wavefunction corresponding to the bare impurity, at the injection momentum.

\subsection{CDWs in the Hartree-Fock approximation}

Understanding the spectral features of polaron formation on top of a charge density wave represents an open problem.
Previous studies have investigated the presence of  umklapp scattering secondary peaks on top of  the RP branch, by using a fully static CDW density~\cite{smolenski2021signatures,shimazaki2021optical}. An exact diagonalization study by two of the authors suggested a crucial role of the quasi-particle band-gap in the spectrum~\cite{amelio2024edpolaron}. Finally,  understanding the role of the CDW phonons and plasmons will be the topic of future research, 
as well as drawing connections to work done in the materials community~\cite{Franchini2021}, where  progress in the ab initio modelling of polarons has been recently achieved~\cite{sio2019abinitio}.

Here, we adopt a drastic approximation and treat the CDW in the Hartree-Fock approximation.
The Hartree-Fock quasi-particles then play the same role as the Bloch quasi-particles. 
While this approach is not able to capture the role of CDW phonons and will fail to describe quantitatively the CDW in the presence of strong quantum fluctuations, we will demonstrate that it qualitatively recovers the features predicted in ED.

More specifically, let us start considering the microscopic fermionic Hamiltonian, in the absence of any external potential:
\begin{equation}
    H^{\rm el} = \sum_{\tilde{\mathbf{k}}}\epsilon_{\tilde{\mathbf{k}}} c_{\tilde{\mathbf{k}}}^\dagger c_{\tilde{\mathbf{k}}}
    +
    \frac{1}{2\mathcal{V}}\sum_{\tilde{\mathbf{k}}\tilde{\mathbf{p}}\tilde{\mathbf{q}}}
    V_{\tilde{\mathbf{q}}}
    c_{\tilde{\mathbf{k}}+\tilde{\mathbf{q}}}^\dagger c_{\tilde{\mathbf{p}}-\tilde{\mathbf{q}}}^\dagger
    c_{\tilde{\mathbf{p}}} c_{\tilde{\mathbf{k}}},
\end{equation}
where, as usual, the $V_{\tilde{\mathbf{q}} = 0}$ 
contribution
is effectively included in the chemical potential and is excluded from the summation.
As fermion-fermion potential, we use the
Coulomb potential $V_\mathbf{q} = \frac{\mathcal{Z}}{|\mathbf{q}|}$~\footnote{Actually, in the numerics a lattice version of the Coulomb potential is used,
$$
    V_\mathbf{q} = 
    \sum_\mathcal{G} \frac{\mathcal{Z}}{|\mathbf{q}+\mathcal{G}|} -  \sum_{\mathcal{\mathcal{G}}\neq 0} \frac{\mathcal{Z}}{|\mathcal{\mathcal{G}}|},
$$
where $\mathcal{G}$ is a reciprocal lattice vector of the microscopic lattice (as opposed to $\mathbf{G}$ in the previous Section, which belonged to the reciprocal lattice of the periodic external potential).
The summation ensures that $V_{\mathbf{q}} = V_{\mathbf{q}+\mathcal{G}}$ (compatible with the fact that wavefunctions are defined only modulo $\mathcal{G}$ on a lattice), and the second term cancels the UV divergence of the first term. Notice that at small $|\mathbf{q}|$ the Coulomb potential is recovered. 
}.
Here, $\mathcal{Z}$ sets the strength of the potential, and, in a real  device,  it will be determined by the length of the lattice constant, which is set to 1 in our units.

\begin{figure*}[t]
    \centering
\includegraphics[width=0.99\textwidth]{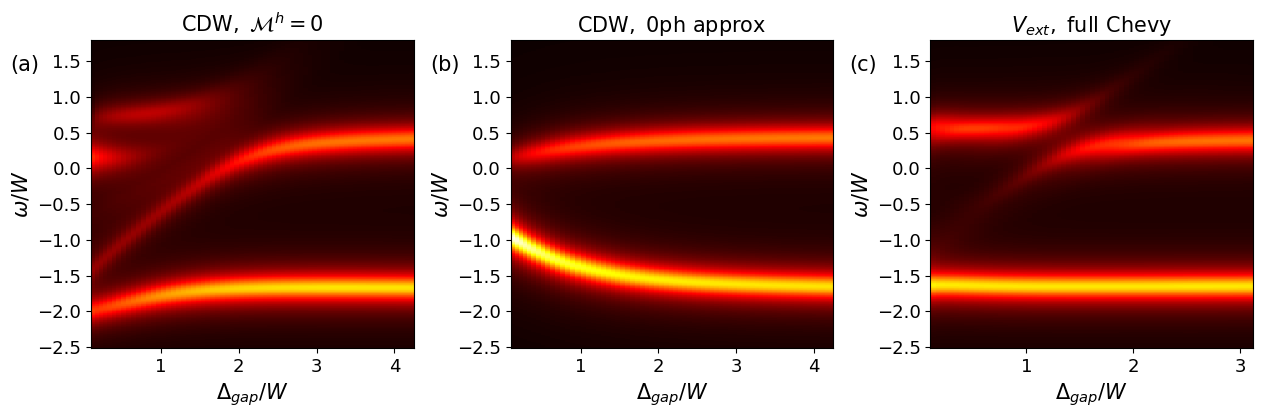}
    \caption{
    Comparative study with respect to Fig.~\ref{fig:ChevyHF}.b. In (a) the hole scattering term is set to 0, while in (b) only the mean-field density of the CDW affects the impurity. In (c) the density modulation is generated by an external potential (no spontaneous CDW), and all Chevy terms are retained. 
    The colorcode is specified in the colorbar of \ref{fig:ChevyHF}.b}
    \label{fig:ChevyHF_comparison}
\end{figure*}

In the Hartree-Fock approximation, one uses the effective quadratic Hamiltonian 
\begin{equation}
    H^{HF} = \sum_\mathbf{k} \epsilon_\mathbf{k} c_\mathbf{k}^\dagger c_\mathbf{k}
    +
   \sum_{\mathbf{k}\mathbf{G}_1\mathbf{G}_2}
    V_{\mathbf{k}}^{HF}(\mathbf{G}_1,\mathbf{G}_2)
    c_{\mathbf{k}+\mathbf{G}_1}^\dagger 
    c_{\mathbf{k}+\mathbf{G}_2}
    \label{eq:H_HF}
\end{equation}
with the nonlocal kernel
\begin{multline}
    V_{\mathbf{k}}^{HF}(\mathbf{G}_1,\mathbf{G}_2)
    =
     \frac{1}{\mathcal{V}}
     \sum_{\mathbf{p}\mathbf{G}}
   (V_{\mathbf{G}_2-\mathbf{G}_1}
    -
    V_{\mathbf{k}-\mathbf{p}-\mathbf{G}}) 
\times
\\
    \times
    \langle
c_{\mathbf{p}+\mathbf{G}+\mathbf{G}_2}^\dagger
c_{\mathbf{p}+\mathbf{G}+\mathbf{G}_1}
    \rangle.
    \label{eq:V_HF}
\end{multline}
This Hamiltonian is simply diagonalized as $
H^{HF}=
    \sum_{\mathbf{k}\alpha}
    E_{\mathbf{k}\alpha} 
    d^\dagger_{\mathbf{k}\alpha} d_{\mathbf{k}\alpha}$,
    where now the $d^\dagger$ operator creates a Hartree-Fock quasi-particle.
A crucial remark is that in a translationally invariant system one would expect 
$\langle
c_{\Tilde{\mathbf{k}}}^\dagger
c_{\Tilde{\mathbf{p}}}
    \rangle =0$ for any $\Tilde{\mathbf{k}} \neq \Tilde{\mathbf{p}}$,
    but here we assume that a CDW is spontaneously formed and reduces translational symmetry, so that 
$\langle
c_{\Tilde{\mathbf{k}}}^\dagger
c_{\Tilde{\mathbf{p}}}
    \rangle \neq 0$ provided that 
    $\Tilde{\mathbf{k}}, \Tilde{\mathbf{p}}$ differ by a reciprocal vector of the CDW.

In practice, the Hartree-Fock ground state can be found iteratively.
We start with an auxiliary external potential, which we turn off in successive iterations. At each step, the Hartree-Fock quadratic Hamiltonian (\ref{eq:H_HF}) (plus the auxiliary potential) is diagonalized, the lowest $N_{\rm v}$ bands are filled, and the Hartree-Fock potential (\ref{eq:V_HF})
is updated. This is repeated until convergence.

\subsection{Numerical results}

We first discuss polaron spectra on top of a CDW. 
We report results for a $9 \times 9$ triangular lattice at one third filling, which is directly relevant to moir\'e TMD systems, but we have verified that we obtain qualitatively similar results for a variety of fillings on both the triangular and square lattices.
For different values of the Coulomb strength $\mathcal{Z}$, we numerically find the Hartree-Fock ground state $|\mathcal{F }\rangle$.
We evaluate the density contrast 
$\Delta n \equiv \max_\mathbf{r}  n(\mathbf{r}) - \min_\mathbf{r}  n(\mathbf{r})$, and the quasi-particle energy gap, $\Delta_{\rm gap} = \min_\mathbf{k} E_{\mathbf{k},N_v+1} - \max_\mathbf{k} E_{\mathbf{k},N_v}$, between the highest valence band and the lowest conduction band. 
The polaron spectrum $A_X(0,\omega)$ at  Bloch momentum $q=0$ is then evaluated 
following the recipe of Eq.~(\ref{eq:A_X}).

The CDW order parameter $\Delta n$ is plotted in Fig.~\ref{fig:ChevyHF}.a as a function of $\Delta_{\rm gap}/W$,
where $W$ is the bandwidth of the non-interacting fermion. The CDW is formed via a first-order phase transition, and the density contrast saturates to 1 for increasing gap, at which point the classical crystal with no quantum fluctuations is achieved.
The inset sketches the triangular density pattern of the CDW.

The polaron spectrum 
$A_X(0,\omega)$
is plotted in Fig.~\ref{fig:ChevyHF}.b, as a function of the quasi-particle gap, and for a two-body binding energy in vacuum of $E_B/W = 0.8$.
For small $\Delta_{\rm gap}/W$, one recovers the usual attractive polaron (AP) and repulsive polaron (RP) branches. Increasing  $\Delta_{\rm gap}/W$, one  observes a line starting slightly above the AP energy, gaining oscillator strength,  and blue-shifting linearly with the gap. At a certain point, this line intersects the RP branch and forms an avoided crossing.
This feature is attributed to a trimer state (on top of the many-body background),
 which generalizes 
the molecule-hole states found in standard Fermi-polaron studies~\cite{parish2011polaron-molecule,parish2023fermi} to the present CDW-background setting.
Indeed, this interpretation can be established by calculating the ground-state energy $E_{\rm t}$
in the Hilbert subspace spanned by states of the form
\begin{equation}
    | \Psi^{\rm t}_{\mathbf{q}} \rangle
    =
\sum_{\mathbf{k}\mathbf{p}\mathbf{G}\alpha\beta} 
\psi_{\mathbf{k}\mathbf{p}}^{\alpha \beta \mathbf{G}} \
d^\dagger_{\mathbf{k}\alpha} d_{\mathbf{p}\beta}
x_{\mathbf{q}+\mathbf{p}-\mathbf{k}+\mathbf{G}}^\dagger
    | \mathcal{F }\rangle.
\end{equation}
Importantly, these molecule-hole states involve the excitation of a Hartree-Fock quasi-particle across the gap 
$\Delta_{\rm gap}$, hence leading to a blueshift of the corresponding energy $E_{\rm t}$
with respect to $\Delta_{\rm gap}$. This behavior is confirmed by the cyan dashed line  displayed in Fig.~\ref{fig:ChevyHF}.b, which nicely captures the spectral function in this regime.

Notice that our Hartree-Fock plus Chevy calculation clarifies the numerical observation of the analogous spectral line found in
the exact diagonalization study \cite{amelio2024edpolaron}. A similar phenomenology has been observed also in theoretical work on twisted heterostructures~\cite{mazza2022strongly}, 
excitonic insulators~\cite{amelio2023polaron}
and Fermi superfluids~\cite{amelio2023two-dimensional}.
Moreover, a redshift of the AP energy is visible for increasing $\Delta_{\rm gap}/W$. This arises from the mass renormalization of the fermions in the CDW, since the quasi-particle bands become more and more flat.

Another interesting feature that is typically present in polaron spectra on top of a CDW is the presence of a umklapp peak above the RP branch.
Our approach is in principle able to capture this effect, originating from the term $\mathcal{M}^{0}$ of Eq.~(\ref{eq:Chevy_CDW_1}).
In Fig.~\ref{fig:ChevyHF}.b no umklapp peak is present though. This is a particular  situation holding for a triangular lattice at one-third filling, and also for a checker-board CDW on top of a square lattice (half-filling).
Indeed, for such  high fillings, it turns out that, in the impurity Hilbert space, there are only two  states 
invariant under translational and inversion symmetry. As a consequence, only two branches can have an overlap with the $q=0$ wave in which the impurity is injected, and be bright.
For lower fillings, we indeed observe umklapp peaks (not shown).

\subsection{The  role of hole scattering processes}

If polaron formation involves only a  very few bands and in the ground state some of these are filled,
the usual phase space  argument for hole scattering being weak does not hold anymore.
The phase space available for the hole is comparable to the one for an electron,
their ratio being just the number of occupied bands over empty ones.

To confirm that hole scattering plays a crucial role, in Fig.~\ref{fig:ChevyHF_comparison}.a we report for comparison the same Chevy computation, but when neglecting the hole scattering term, i.e. $\mathcal{M}^{\rm h} \to 0$.
A strong qualitative discrepancy with respect to Fig.~\ref{fig:ChevyHF}.b  is present  at small $\Delta_{\rm gap}$. 
First, the AP line is sizably lower in energy in Fig.~\ref{fig:ChevyHF_comparison}.a, since hole scattering processes correspond to a repulsive hole-impurity interaction. Neglecting this repulsion lowers the energy.
Moreover, an extra line is visible when hole scattering is neglected. As analyzed in Appendix \ref{app:B}, the wavefunction of the state associated with this extra peak reveals correlation between the impurity and the ground-state fermionic densities. This is also typical of the AP, while 
anti-correlation is expected for the RP. 
Since the hole density naturally correlates with the ground-state fermionic density, we speculate that the extra peak originates from the absence of hole repulsion, in a loose analogy with a weakly-bound state that unbinds when introducing repulsive interaction.

At large $\Delta_{\rm gap} \gg |g|$, instead, the creation of a quasi-particle-hole pair excitation is very off-resonant and does not play any role. In other words, at large $\Delta_{\rm gap}$ the polaron spectrum is purely due to the scattering of the impurity with a static potential
generated by the extremely rigid CDW.
This is demonstrated in Fig.~\ref{fig:ChevyHF_comparison}.b,
where the spectrum is computed by setting $\mathcal{M}^{\rm ph},\mathcal{M}^{\rm p},\mathcal{M}^{\rm h} \to 0$, and  the term $\mathcal{M}^{\rm 0}$ plays the main role. 
In other words, this is analogous to solving in the zero particle-hole excitations (0ph) subspace
\begin{equation}
    | \Psi^{0ph}_{\mathbf{q}} \rangle
    =
\sum_{\mathbf{G}} \psi_{\mathbf{G}} x^\dagger_{\mathbf{q}+\mathbf{G}} 
 | \mathcal{F }\rangle.
\end{equation}
We call this mean-field treatment, where the impurity scatters with the Hartree potential of the CDW,
the 0ph approximation (we choose this notation not to confuse with the purely fermionic Hartree-Fock  mean-field).
These considerations also entail that at large $\Delta_{\rm gap}$ the RP branch is just a $\delta$-like line, without the broadening present in a gapless Fermi sea.

Finally, in Fig.~\ref{fig:ChevyHF_comparison}.c
we study the case where the fermions are not interacting, but an external potential is present and induces a periodic density modulation of the fermions.
In this case, we find that the polaron spectrum at the Chevy level presents the same qualitative features as for the CDW of Fig.~\ref{fig:ChevyHF}.b, including the  avoided crossing in the RP branch related to the quasi-particle gap.
% We attribute the smaller avoided crossing observed 
% Fig.~\ref{fig:ChevyHF_comparison}.c as compared to 
% Fig.~\ref{fig:ChevyHF}.b to the fact that, for equal gap $\Delta_{\rm gap}$, the quasi-particle bands arising from $V^{\rm ext}$ turn out to be flatter than the ones arising from the HF spontaneously formed CDW (correspondingly, $\Delta n$ is larger in the first case). This entails less spatial overlap, hence smaller avoided crossing, between the RP and trimer states, which are predominantly localized on the low and high density sites, respectively. 

~

Let us summarize the findings of this Section. We have first introduced the Chevy-ansatz formalism for a spatially periodic background, and we have computed the polaron spectra for band insulators and CDWs, which, at the Hartree-Fock level, are described in terms of quasi-particle bands. An avoided crossing in the RP branch reflects the magnitude of the quasi-particle gap in the background medium, and the AP energy is shown to redshift due to the mass renormalization and localization of the fermions.
Neglecting hole-scattering processes, which are of repulsive nature, results in a spurious peak and low AP energies. In the regime of large quasi-particle gap, the fermionic bath is incompressible and neglecting the dressing by particle-hole excitations  (i.e.~the 0ph approximation) is justified.

\section{Polarons across the Mott transition}
\label{sec:mott}

\begin{figure*}[t]
    \centering
\includegraphics[width=0.99\textwidth]{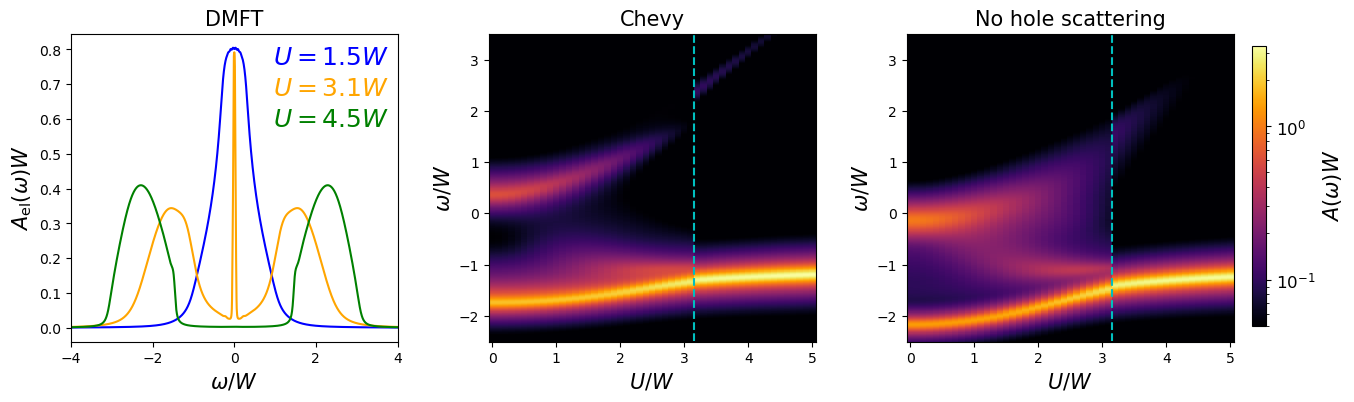}
    \caption{
    (a)
    Phenomenological electronic density of states 
    $A_{\rm el}(\omega)$ for three different $U$'s, in the metallic phase (blue line), just below the Mott transition (orange), and deep into the insulating regime (green).
    (b) 
    Polaron spectrum across the Mott transition in a Chevy calculation with infinite impurity mass. The vertical cyan dotted indicates the Mott transition.
    A logarithmic scale is used.
    (c) Same as in (b), but having neglected the hole-impurity scattering processes.
    }
    \label{fig:Mott_heavy}
\end{figure*}

Besides trivial band insulators and CDWs, a paradigmatic type of insulator is provided by the Mott state~\cite{Gebhard1997}.
This insulating phase of matter arises from strong  repulsion between the fermions which, in half-filled systems, 
leads to the localization of charges in the absence of symmetry breaking.

Here, we  are interested 
in the evolution of polaron spectra 
when the underlying fermionic system undergoes a Mott transition.
Specifically, we describe the dynamics of the  bath by using the Fermi-Hubbard Hamiltonian
\begin{equation}
H^{\rm el}
=
-t \sum_{i j\sigma}
c_{i\sigma}^\dagger c^{\phantom{\dagger}}_{j\sigma} +
U \sum_{i} c^{\dagger}_{i \uparrow} c^{\phantom{\dagger}}_{i \uparrow}
c^{\dagger}_{i \downarrow} c^{\phantom{\dagger}}_{i \downarrow},
\label{eq:FermiHubbardH}
\end{equation}
where $t$ is the hopping amplitude between sites $i$ and $j$, $U$ is the strength of the 
local repulsion, and $\sigma=\uparrow,\downarrow$ is the electron spin.
We compute polaron spectra using the Chevy and the  infinite mass approximations. 
As shown in the Appendix~\ref{app:A}, we first observe that, at the Chevy level and for a heavy impurity in a background of quasi-particles, only the momentum-averaged electronic density of states
$A_{\rm el}(\omega)$ enters the calculation of the polaron spectrum.
Here, we compute the electronic density  of states 
$A_{el}(\omega)$
by means of  
Dynamical Mean-Field Theory (DMFT) using the Iterated Perturbation Theory (IPT) solver~\cite{georges1996rmp}.
Specifically, we consider the Bethe lattice with bandwitdh $W$.
The electronic density of states $A_{\rm el}(\omega)$ 
at half filling
is plotted in Fig.~\ref{fig:Mott_heavy}.a for different values of the electron-electron interaction over the bandwidth $U/W$. The Mott transition occurs approximately
at the critical interaction strength $U_c/W \simeq 3.15$.
In the metallic phase at $U<U_c$, the density of states has a characteristic three 
 peaks structure.
The central band is metallic and it describes coherent quasi-particles close to the Fermi level.
The quasi-particle peak looses spectral weight as 
approaching the Mott transition, whereas the 
 density of states at the Fermi level stays constant, as required by the Luttinger theorem~\cite{luttinger1960}.
 The two side bands describe incoherent excitation at energies $\pm U/2$ and describe 
 the formation of doubly occupied (or empty) sites. 
 At the Mott transition, the coherent peak disappears and the spectral density 
 is characterized by the two Hubbard bands of width $~W/2.$

We now immerse the heavy impurity in the fermionic bath.
We can work in a Lee-Low-Pines frame~\cite{lee1953}, where the impurity is confined in a specific lattice site, with creation operator $x^\dagger$. 
The heavy impurity condition entails that the effective electron-electron recoil term  is negligible, so that the electronic Hamiltonian  is the one of Eq.~(\ref{eq:FermiHubbardH}) also in this frame.
Also, we assume that the impurity interacts with strength
$g$ with only the spin up fermions.

As sketched in the Appendix \ref{app:A}, if one assumes the Chevy approximation and neglects fermionic vertices (apart from those already included in the DMFT  Green's function $A_{\rm el}$),
the polaron self-energy in the Chevy approximation can
be computed by solving 
the auxiliary Hamiltonian
\begin{equation}
    H^{\rm aux}
    = \epsilon_X  x^\dagger x 
    +
    \sum_{j=1}^{N_\omega} \omega_j c^\dagger_j c_j 
    +
    \frac{g}{\mathcal{V}} \sum_{jl} g_{jl} c^\dagger_j c_l x^\dagger x,  
    \label{eq:H_Mott_polaron}
\end{equation}
where
$N_\omega$ is the number of representative electronic states, the effective couplings are $g_{jl} = \frac{g}{N_\omega} \sqrt{Z_j Z_l}$ and
 $\omega_j, Z_j$
are chosen according to the constraint
$\lim_{N_{\omega} \to \infty} \frac{1}{N_{\omega}} \sum_j Z_j \delta(\omega -\omega_j) = A_{\rm el}(\omega)$. Clearly, many  choices are possible for $\omega_j$ and $Z_j$, and, in practice, we use a grid of 
$N_{\omega} \sim 150$ points, denser around the Fermi level.

At this point, we use the usual Chevy recipe, where the fermionic ground state $|\mathcal{F}\rangle$ is obtained by filling the electronic states below the Fermi level. For clarity, we rewrite the Chevy ansatz in this context
\begin{equation}
    |\Psi\rangle 
    =
    \left\{ \psi_0 +
    \sum_{jl} \psi_{jl} c^\dagger_j c_l
    \right\}  x^\dagger |\mathcal{F}\rangle.
\end{equation}

We remark that one could have avoided the heavy impurity approximation, and introduced for every electronic momentum an $A_{\rm el}(\mathbf{k}, \omega)$ density of states, represented by a set of discrete states.
This would have been quite heavy computationally, since, for each momentum,
one should have introduced many states to described the broad, incoherent Hubbard  peaks.
In principle, one could also add interactions between the auxiliary quasi-particles $c_j$, similarly to what done in \cite{yi2015polarons,amelio2023polaron,amelio2023two-dimensional} for  a BCS condensate background; this direction is left for future investigations.

Numerically solving the polaron spectra associated with the Hamiltonian (\ref{eq:H_Mott_polaron})
yields the spectrum shown in Fig. \ref{fig:Mott_heavy}.b,
for $g/W = -2$.
At small $U$, the metallic nature of the electronic bath leads to  conventional AP and RP branches.
At large $U$, conversely, the Mott insulator is completely incompressible, and the bath just redshifts the impurity by $g/2$, the electronic density being $1/2$. 
The RP branch is barely visible, since the 0ph and 1ph sectors are quite off-resonant, and very weakly hybridized. 
This is consistent with findings of the previous Section and with the second-order perturbation theory prediction in the Bose-Hubbard Mott transition~\cite{colussi2023lattice}.

At intermediate $U$'s,
there is a  cusp behavior of  the AP  branch at the Mott transition,  indicated by the dotted cyan line.
The RP branch also shows a small jump.
Moreover, a fainter line is present at intermediate energies, and it gains oscillator strength close to the metallic side of the transition, and disappears in the insulating region. 
Inspection of the wavefunctions of the states contributing to this line shows that 
they consist of particle-hole excitations living in the narrow metallic band.
Therefore, this feature highlights the fact that in the fermionic bath, close to the Mott transition, there exists two types of competing excitations, i.e. the quasi-particles close to the Fermi level and the incoherent Hubbard bands, that contribute to the formation of the bound state. 
This is analogous to the splitting of the RP branch observed in heterostructures undergoing the doping driven Mott localization~\cite{mazza2022strongly}.

In Fig. \ref{fig:Mott_heavy}.c, 
we report the same calculation, but having removed the hole-impurity scattering processes.
Even though there are no dramatic qualitative differences, there is a sizable quantitative discrepancy in the position of the peaks with respect to the full calculation.
This is expected because of the comparable particle and hole phase spaces for scattering.
In particular, neglecting the impurity-hole repulsion lowers the AP energy.

A final comment is in order and relates to the metallic phase.
For a heavy impurity, one would expect Anderson orthogonality catastrophe and a power-law broadening of the AP line in the metallic phase~\cite{schmidt2018}. This behavior is not captured by Chevy ansatz, but this analysis goes beyond  our goals. One  could
in principle solve the effective Hamiltonian (\ref{eq:H_Mott_polaron}) exactly using the functional renormalization group method, which takes into account any order of particle-hole excitations, but for the sake of simplicity we do not discuss this here.

\begin{figure}[t]
    \centering
\includegraphics[width=0.32\textwidth]{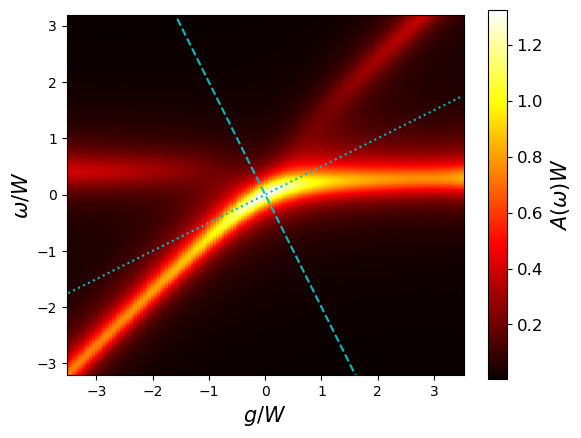}
    \caption{
    Polaron spectrum as a function of $g$ and for $U/W=0.5$. The cyan dotted line represents the mean-field shift 0ph shift of the impurity.
    The particle-hole symmetry of the electronic background results in a reflection symmetry across the cyan dashed line, which is orthogonal to the dotted line.
    }
    \label{fig:PHsymmetry}
\end{figure}

\subsection{Particle-hole duality}

The electronic density of states 
plotted in Fig.~\ref{fig:Mott_heavy}.a is symmetrical across the Fermi level,
and leads to the particle-hole duality we now describe  in rather general terms.

We introduce the particle-hole transformation
$h_j = c_{-j}^\dagger$,
where the notation $-j$ denotes the index $N_\omega-j+1$, satisfying
$\omega_{j} = - \omega_{-j}, Z_j = Z_{-j}$
~\footnote{While this argument is quite general, for definiteness we explain it in the context of the Hamiltonian (\ref{eq:H_Mott_polaron}). In particular, we assume that our grid of auxiliary fermionic states respects the particle-hole symmetry, i.e. it is symmetric across the Fermi level.}.
Applying this transformation to the Hamiltonian (\ref{eq:H_Mott_polaron}),
using the anti-commutation relations and redefining the summation indices $j \to -j$, we obtain
\begin{equation}
    H 
    = (\epsilon^X + g)  x^\dagger x 
    +
    \sum_{j=1}^{N_\omega} \omega_j h^\dagger_j h_j 
    -
    \frac{g}{\mathcal{V}} \sum_{j,k} \sqrt{Z_j Z_k} h^\dagger_j h_k x^\dagger x.  
    \label{eq:Hhole}
\end{equation}
When the fermionic degrees of freedom are traced out, this establishes a relation between the impurity spectrum associated with the original Hamiltonian and the one associated with Eq.~(\ref{eq:Hhole}).
This has the same form as the original Hamiltonian, but for a shift $g$ of the impurity energy and for the fact that the interaction term comes with an extra minus sign.

This entails a geometric symmetry in the polaron spectrum, as shown in Fig.~\ref{fig:PHsymmetry} for $U/W=0.5$.
The positive and negative $g$ sides are clearly related. More precisely, the cyan dotted line denotes the mean-field shift $g/2$ (the $1/2$ is due to being at half-filling), and the spectrum has a reflection symmetry across the cyan dashed line, which is the normal to the mean-field shift passing through the origin.
Moreover, two branches are visible on both the  positive and negative $g$ side.
 This duality also applies to
 other  particle-symmetric baths, such as the chiral edge states or the Haldane model at half filling~\cite{vashisht2024chiral}. 
On the contrary, for a conventional Fermi polaron in a weakly doped band, one would have just one nearly flat line on the repulsive $g>0$ side.

~

Summarizing, this Section 
investigated the polaron spectroscopy of a Fermi-Hubbard system at half-filling. The application of the Chevy-ansatz method was made possible by treating the background fermionic medium by introducing free auxiliary fermions that mimic the Green's function calculated from DMFT. Importantly, the polaron spectrum is shown to exhibit a sharp qualitative change at the Mott transition. Besides, the peculiar duality arising from particle-hole symmetry was shown to lead to characteristic symmetries in the polaron spectrum. Finally, neglecting hole-impurity scattering is shown to quantitatively alter the spectral lines.

\section{Conclusions}
\label{sec:conclusions}

In summary, we extended the Chevy-ansatz framework to the context of
insulating states of fermionic matter.
As a notable result, we have demonstrated that the polaron spectrum of a mobile impurity immersed in a  CDW displays a clear signature of the underlying quasi-particle band gap. Then, using input from DMFT, we have shown that the polaron spectrum of an impurity injected in the strongly-interacting Fermi-Hubbard model exhibits a sharp signature at the Mott transition. Altogether, these results indicate how polaron spectroscopy can be used as a probe of insulating fermionic matter.

We have emphasized the key role of hole-impurity repulsion, which is relevant in this context due to the comparable size of particle and hole scattering phase spaces:~neglecting these processes can indeed result in low attractive-polaron energies, and to additional spectral lines.
In the case of the half-filled Fermi-Hubbard model, the particle-hole symmetry of the fermionic background was shown to lead to a duality between positive and negative couplings and to striking symmetries in the resulting impurity spectral function.

An important remark is in order concerning the observability of the avoided crossing shown in Fig.~\ref{fig:ChevyHF}.b 
in TMD experiments. In TMD experiments, the trion binding energy (which corresponds to the impurity-fermion 2-body binding energy in vacuum) is not straightforwardly tunable, and it is typically around 25 meV.
This is larger than typical electronic gaps, so that such an avoided crossing is yet to be observed.
Also, polaron formation involves many empty bands, and hole scattering is expected to be negligible. 
Hopefully, future experiments will achieve ways to tune the trion binding energy or obtain larger electronic gaps.

On the theory side, it will be interesting to investigate the role of CDW phonons and plasmons, as well as the inclusion of fermionic vertices in the treatment of polaron formation across the Mott transition. The reliability of the Chevy approximation will then need to be benchmarked against numerical methods, which at present are also being actively developed~\cite{vashisht2024chiral}.
Our approach could also be adapted to study bosonic supersolids, where 
Bloch theorem entails the existence of many Bogoliubov bands, rather than quasi-particle bands~\cite{simons2024polaronsupersolid}.
Another interesting direction concerns addressing  polaron formation in periodic external potentials by truncating the number of bands  and introducing a renormalized interaction~\cite{buchler2010microscopic};
we note that the case where the impurity also feels an external potential is relevant for excitons in moir\'e TMD bilayers~\cite{Naik2022,polovnikov2022coulombcorrelated,kiper2024confinedtrionsmottwignerstates}. 
Finally, while here we considered the impurity as a point-like distinguishable particle, it would be very interesting to include the microscopic modelling of the exciton, which is also expected to be affected by the electronic correlations of the bath, in particular in the Mott phase~\cite{huang2023mott,huang2023spin,kiper2024confinedtrionsmottwignerstates}.

\section*{Acknowledgements}
We are grateful to Atac Imamoglu, Eugene Demler, Haydn Adlong, 
Alessio Recati, Amit Vashisht, Oriana K. Diessel and Adriano Amaricci for useful discussions.
This research was financially supported by the ERC grant LATIS, the EOS project CHEQS and the FRS-FNRS (Belgium). 
G.M. acknowledges support from the MUR - Italian Minister of University and Research - under the ``Rita Levi-Montalcini'' program.
Computational resources have been provided by the Consortium des \'Equipements de Calcul Intensif (C\'ECI), funded by the Fonds de la Recherche Scientifique de Belgique (F.R.S.-FNRS) under Grant No. 2.5020.11 and by the Walloon Region.

\appendix

\section{Diagrammatic remarks}
\label{app:A}

\begin{figure*}[t]
    \centering
\includegraphics[width=0.88\textwidth]{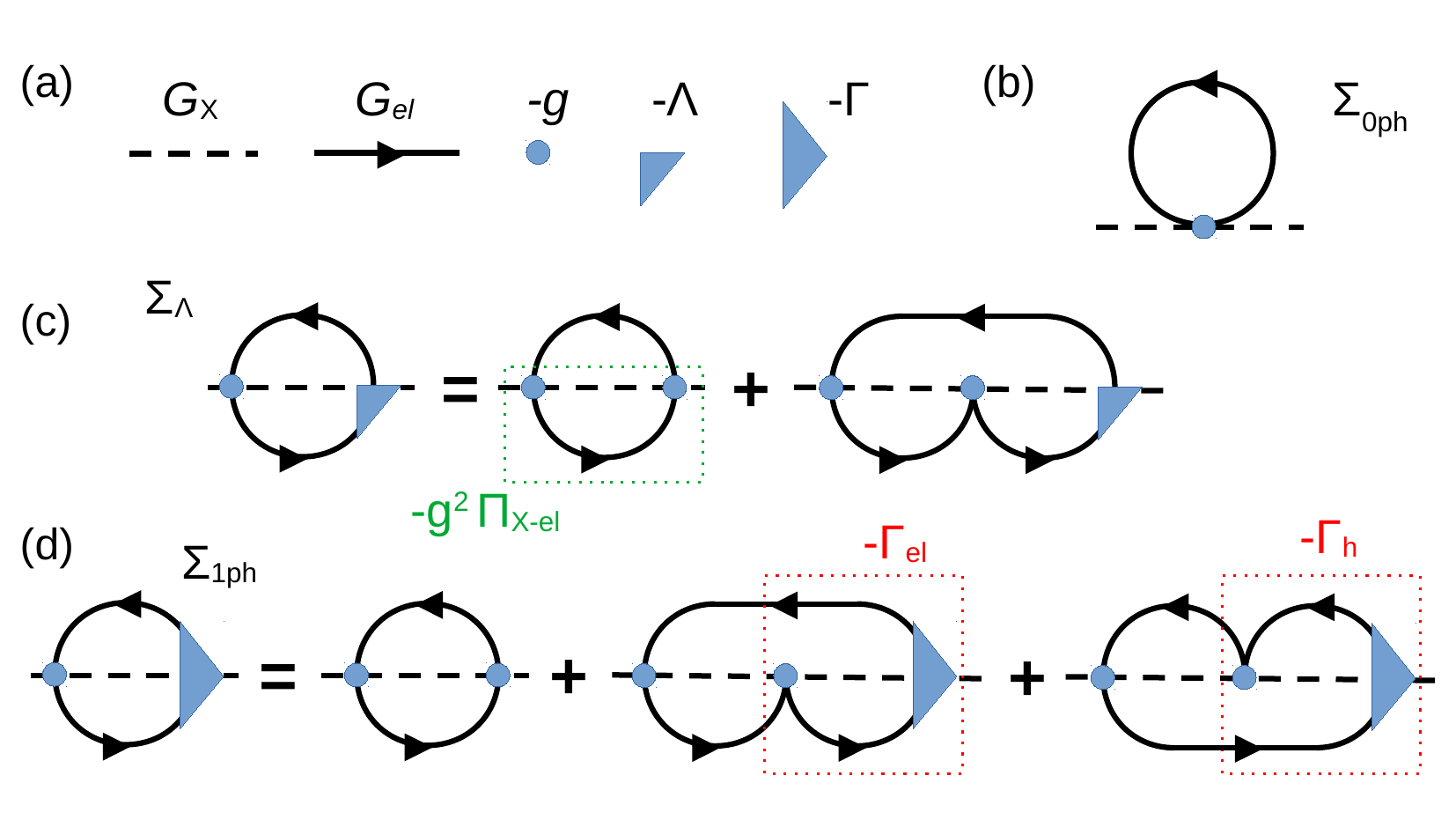}
    \caption{
    Diagrammatic resummation of Chevy diagrams. 
    (a) Diagrammatic representation of the Green's functions and vertices.
    (b) Mean-field term of the self-energy $\Sigma_{0ph}$.
    (c) Self-energy from diagrams with a single particle-hole excitation $\Sigma_{\Lambda}$
    in the ladder approximation (i.e. only impurity-electron processes are resummed). The green dotted rectangle hints at the impurity-electron bubble.
    (d) Self-energy from diagrams with a single particle-hole $\Sigma_{\rm 1ph}$,
    but resumming also the hole-impurity scattering processes. 
    The red dotted rectangles highlight the decomposition of the vertex according to 
    $\Gamma = g + \Gamma_{el} + \Gamma_{h}$.
    }
    \label{fig:diagrams}
\end{figure*}

In this Section, we want to illustrate the diagrammatic resummation of the T-matrix (or ladder) and Chevy approximations, for an electronic bath representable by non-interacting quasi-particles. 
In particular, we show that the 
the Chevy diagrams are also resummable, but lead to a self-consistent equation which is difficult to invert.
The goal of this Section is thus mainly conceptual.

The Green's function of the impurity is
$G_X(Q) = \frac{1}{\Omega - \epsilon^X_{\mathbf{Q}} - \Sigma_X(Q)}$,
with the
self-energy  being
\begin{equation}
    \Sigma_X(Q)
    = 
    \Sigma_{\rm 0ph}(Q) +
    \Sigma_{\rm 1ph}(Q) + ...,
\end{equation}
where the truncation refers to processes containing more than one particle-hole excitations.
Notice that an energy-momentum notation is used: $Q = (\mathbf{Q},i\Omega_n)$, where $i\Omega_n$ is a Matsubara frequency.
In the following, for simplicity we limit ourselves to a translationally invariant system, 
for which the mean-field polaron self-energy is just  $\Sigma_{\rm 0ph}(Q) = g n$.

In Fig.~\ref{fig:diagrams}.a we recall the diagrammatic representation of the propagator and vertices.
The mean-field term of the self-energy $\Sigma_{\rm 0ph}(Q)$ is depicted in Fig.~\ref{fig:diagrams}.b,
while $\Sigma_{\rm 1ph} \simeq \Sigma_{\Lambda}$ in the ladder approximation is illustrated in Fig.~\ref{fig:diagrams}.c. 
In Fig.~\ref{fig:diagrams}.d,  hole scattering processes are included in the full $\Sigma_{\rm 1ph}$.

We remark that no fermion-fermion vertex is considered in the diagrams of Fig.~\ref{fig:diagrams}. This can be a very bad approximation, in particular whenever the quasi-particles are supposed to be paired up into excitons or Cooper pairs \cite{amelio2023polaron}.
When this approximation is assumed, 
we now show that the polaron self-energy 
can be computed representing
the electronic bath  in terms of non-interacting quasi-particles,
allowing for a Hamiltonian treatment like in Eq.~(\ref{eq:H_Mott_polaron}).
The idea is that, 
according to the Kramers-Kronig, the electron Green's function can always be written as
 $G_{\rm el}(\mathbf{k},z) = \int d\omega \frac{A_{\rm el}(\mathbf{k},\omega)}{z-\omega}$, with 
$A_{\rm el}(\mathbf{k},\omega) = - \frac{1}{\pi}{\rm Im} G_{\rm el}(\mathbf{k},\omega+i0^+)$.

In general, $A_{\rm el}(k)$ can contain broad peaks, and this is true for example in the case of the Fermi-Hubbard model.  
Nonetheless, $A_{\rm el}(k)$ can be replaced by $N_{\omega}$ sharp peaks $A_{\rm el}(\mathbf{k},\omega) \sim \frac{1}{N_{\omega}} \sum_j Z_j(\mathbf{k}) \delta(\omega -\omega_j(\mathbf{k}))$, where the approximation becomes exact  in the limit $N_{\omega} \to \infty$.

Let's now consider for definiteness the second-order Feynmann diagram appearing in Fig.~\ref{fig:diagrams}.(c), but the same reasoning is straightforwardly adapted to a generic  diagram contributing to the impurity self-energy.
This contribution reads
\begin{multline}
    \Sigma^{(2)}_X(Q) =
    -\frac{g^2}{\beta^2 \mathcal{V}^2}
    \sum_{p,k}
     G_{\rm el}(p) G_{\rm el}(k)
     G_{\rm X}(Q-k+p) = 
     \\
     = -\sum_{p,k,j,l} \frac{g_{jl}(\mathbf{k},\mathbf{p}) g_{lj}(\mathbf{p},\mathbf{k})}{\beta^2 \mathcal{V}^2} G_j(p) G_l(k)
     G_{\rm X}(Q-k+p),
\end{multline}
where in the second step we introduced the free fermion propagator
$G_j (\omega, \mathbf{k}) = \frac{1}{\omega - \omega_j(\mathbf{k})} $
associated to the auxiliary fermionic mode $j$ and the couplings
$g_{jl}(\mathbf{k},\mathbf{p})
=
\frac{g}{N_\omega}\sqrt{Z_j(\mathbf{k}) Z_l(\mathbf{p})}$.
A similar rearrangement can be performed for any diagram of the polaron self-energy, since every fermionic propagator has the vertex $g$ at both endpoints. 
On the other hand, the same diagrammatic expansion follows from the Hamiltonian
\begin{multline}
    H^{\rm aux}
    =
    \sum_{\mathbf{k}j} 
    \omega_j(\mathbf{k}) c_{\mathbf{k}j}^\dagger c_{\mathbf{k}j}
    +
    \sum_{\mathbf{q}} \epsilon_X({\mathbf{q}})
    x_{\mathbf{q}}^\dagger x_{\mathbf{q}}
    +
    \\
    +
    \frac{1}{\mathcal{V}}
    \sum_{\mathbf{k}\mathbf{p}\mathbf{q}jl}
    g_{jl}(\mathbf{k},\mathbf{p})
    x_{\mathbf{q}+\mathbf{k}-\mathbf{p}}^\dagger x_{\mathbf{q}}
    c_{\mathbf{k}j}^\dagger c_{\mathbf{p}l}
    \label{eq:H_eff}
\end{multline}
entailing that one can diagonalize $H_{\rm eff}$
in order to compute the polaron spectrum.
Once again, the huge approximation required is to have neglected fermionic vertices.

The auxiliary Hamiltonian of Eq.~(\ref{eq:H_Mott_polaron})
holds in the infinite impurity mass limit, by performing the Lee-Low-Pines transformation and 
using the strategy explained here, noticing that the momentum indices can be dropped.

\subsubsection*{Ladder approximation}

In the ladder approximation, injection of the impurity creates a particle-hole pair, but at that point the impurity only scatters with the particle. This is a reasonable approximation, if the Fermi sea is small and the phase space of the hole is quite limited.
In this approximation, one has 
\begin{equation}
    \Sigma_{\rm 1ph}(Q)
    \simeq
     \Sigma_{\rm \Lambda}(Q)
    = 
    \frac{1}{\beta \mathcal{V}}
    \sum_{\mathbf{p},ip_n}
     G_{\rm el}(p)
    \Lambda(Q+p)
\end{equation}

The two body T-matrix (here taken without first order term in $g$, in contrast to $\Tilde{\Lambda}$ in Eq.~(\ref{eq:tilde_Lambda})) is given in the ladder approximation by
\begin{equation}
    \Lambda(Q) = g \frac{g \Pi_{\rm X-el}(Q)}{1 - g \Pi_{\rm X-el}(Q)}
\end{equation}
with (minus) the fermion-impurity bubble diagram 
being
\begin{equation}
    \Pi_{\rm X-el}(Q)
    = - \frac{1}{\beta \mathcal{V}} \sum_{\mathbf{k},ik_n}
    G_X(Q-k) G_{\rm el}(k).
\end{equation}

\begin{figure*}[t]
    \centering
\includegraphics[width=0.88\textwidth]{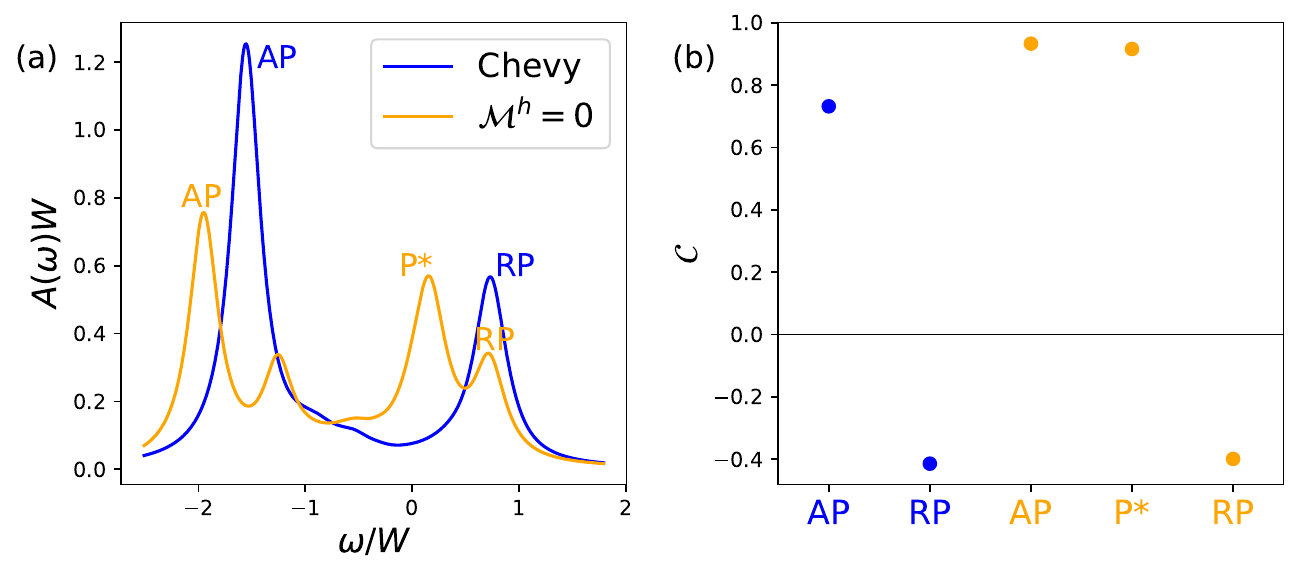}
    \caption{
    (a) Slices  at $\Delta_{\rm gap} \simeq 0.3 W$ of the spectral function from Fig.~\ref{fig:ChevyHF}.b and Fig.~\ref{fig:ChevyHF_comparison}.a, for a polaron in a CDW. The blue curve is the full Chevy calculation, while the orange line is without hole scattering processes. The spurious peak is labelled $P^*$.
    (b) The exciton density-fermion density correlator $C$ is evaluated in the states which contribute more to each peak.
    }
    \label{fig:extra_peak}
\end{figure*}

For quasi-particles
with dispersion $\xi_{\mathbf{k}}$, the self-energy  simplifies to
\begin{equation}
    \Sigma_X(\mathbf{Q}, \Omega)
    = 
    g n + 
    \frac{1}{A}
    \sum_{\mathbf{p}} 
   n_F(\xi_{\mathbf{p}}) 
    \Lambda(\mathbf{Q} + \mathbf{p},  \Omega + \xi_{\mathbf{p}}),
\end{equation}
which basically recovers Eq.~(\ref{eq:SigmaX_simple}), since $ \Tilde{\Lambda} = g + \Lambda$.

\subsubsection*{Chevy resummation}

As visible in Fig. \ref{fig:diagrams}.d, the one-particle-hole excitation diagrams including also hole scattering processes are also resummable.
We call $\Gamma(Q,k,p)$ the Chevy vertex of the theory, so that the self-energy within this approximation reads
\begin{equation}
    \Sigma_{\rm 1ph}(Q) =
    \frac{-g}{\beta^2 \mathcal{V}^2}
    \sum_{p,k}
     G_{\rm el}(p) G_{\rm el}(k)
     G_{\rm X}(Q-k+p)
    \Gamma(Q,k,p), 
\end{equation}
where the minus sign comes from the fermionic loop.

The Feynmann rules yields a self-consistent equation for $\Gamma$:
\begin{equation}
    \Gamma(Q,k,p)
    = g
+
    \Gamma_{\rm el}(Q,p) 
    +
     \Gamma_{\rm h}(Q,k) 
\end{equation}
where $\Gamma_{\rm el}$ gathers all the diagrams which start with an impurity-electron scattering process, while in $\Gamma_{\rm h}$ an impurity-hole scattering occurs first.
We remark that, while for a generic translationally invariant theory $\Gamma(Q,k,p)$
depends on three energy-momenta, by further assuming  contact instantaneous interactions this complex dependency simplifies a lot, since  $\Gamma_{\rm el}(Q,p)$ does not depend on the electron energy-momenta, and similarly  $\Gamma_{\rm h}(Q,k)$ does not depend on $p$.
This is seen from the explicit expressions
\begin{equation}
   - \Gamma_{\rm el}(Q,p)
    =
     \frac{g}{\beta V}
    \sum_{k'}
     G_{\rm el}(k')
     G_{\rm X}(Q-k'+p)
    \Gamma(Q,k',p),
\end{equation}
\begin{equation}
   - \Gamma_{\rm h}(Q,k)
    =
     \frac{g}{\beta V}
    \sum_{p'}
     G_{\rm el}(p')
     G_{\rm X}(Q-k+p')
    \Gamma(Q,k,p'). 
\end{equation}

Inverting these equations is in practice cumbersome,
and the strategy we adopted in this paper is to
rather use the effective quasi-particle Hamiltonian approach of Eq.~(\ref{eq:H_eff}).

\section{Nature of the extra peak}
\label{app:B}

In this Appendix we give some insight on the nature of the spurious peak which is observed in Fig.~\ref{fig:ChevyHF_comparison}.a when the hole scattering processes are neglected.
First of all, we replot in Fig.~\ref{fig:extra_peak}.a
the slice of the spectral function
for a relevant value of quasi-particle gap
$\Delta_{\rm gap} \simeq 0.3 W$ and density contrast $\Delta n \simeq 0.5$,
comparing the full Chevy calculation 
in blue with the one in red, obtained by leaving out hole scattering, i.e. by setting $\mathcal{M}^h=0$. 
In the latter case, the AP energy is significantly lower, the trimer-like peak is more pronounced, and, above all
there is an extra peak, labelled $P^*$ in the plot.
Below we show that the states contributing to this peak behave more like the AP than the RP.

To this end, we quantify the correlations between the exciton density 
$n^X_{\mathbf{r}} = 
\langle \psi | x_{\mathbf{r}}^\dagger x_{\mathbf{r}}
|\psi\rangle
$
in a given state $|\psi\rangle$ and the fermionic density
$n_{\mathbf{r}} = 
\langle \mathcal{F} | c_{\mathbf{r}}^\dagger c_{\mathbf{r}}
|\mathcal{F}\rangle
$
in the ground state without impurity  $|\mathcal{F}\rangle$ by considering the quantity
\begin{equation}
    \mathcal{C} =
    \frac{\overline{ n^X_{\mathbf{r}}
    n_{\mathbf{r}}}}
    {\overline{ n^X_{\mathbf{r}}
    } \cdot \overline{ 
    n_{\mathbf{r}}^{~}} } - 1,
\end{equation}
where the bar denotes spatial average.
We compute $\mathcal{C}$
by considering for each peak the state 
 $|\psi\rangle$
 with the largest oscillator strength contributing to that peak. 
 The results are reported in 
Fig.~\ref{fig:extra_peak}.b,
where the colors and labels are in correspondence with panel (a).
As expected, the AP states display positive correlation: in the many-body ground state, the impurity tends to be attracted and possibly bind to regions of high fermionic density.
Conversely, the RP show anti-correlation, consistently with the picture of the impurity scattering states avoiding the fermions. 
Interestingly, the extra peak also shows strong positive correlation.
Considering that in general hole formation requires and correlates with the fermionic density, it is reasonable that the impurity-hole repulsion washes out this peak (similarly to the AP peak being pushed up in energy). 

\bibliography{bibl}

\end{document}